\documentclass[acmtog]{acmart}
\acmSubmissionID{207}

\usepackage{booktabs} 
\usepackage{xcolor}
\usepackage{layouts}
\usepackage{enumitem}
\usepackage{multirow}
\usepackage{nicematrix}
\usepackage{tabularx}
\usepackage{wrapfig}
\usepackage{colortbl}
\usepackage[normalem]{ulem}

\usepackage[ruled]{algorithm2e} 

\SetAlFnt{\small}
\SetAlCapFnt{\small}
\SetAlCapNameFnt{\small}
\SetAlCapHSkip{0pt}

\setcopyright{acmlicensed}
\acmJournal{TOG}
\acmYear{2023} 
\acmVolume{42} 
\acmNumber{6} 
\acmArticle{229} 
\acmMonth{12} 
\acmPrice{15.00}
\acmDOI{10.1145/3618316}

\newif\ifshownotes
\shownotestrue 

\ifdefined\MakeWithNotes
  \shownotestrue
\fi
\ifdefined\MakeWithoutNotes
  \shownotesfalse
\fi

\ifshownotes
  \newcommand{\colornote}[3]{{\color{#1}\bf{#2: #3}\normalfont}}
  \newcommand{\colornoteTwo}[3]{{\color{#1}\bf{#3}\normalfont}}
  \newcommand{\colornoteThree}[2]{{\color{#1}\bf{#2}\normalfont}}      
\else
  \newcommand{\colornote}[3]{}
  \newcommand{\colornoteTwo}[3]{}
  \newcommand{\colornoteThree}[2]{}      
\fi

\definecolor{darkgreen}{rgb}{0.0,0.65,0}

\newcommand{\etal}{\textit{et al}.}

\SetKwComment{Comment}{/* }{ */}
\SetKwRepeat{Do}{do}{while}

\begin{document}
\title{Editing Motion Graphics Video via Motion Vectorization and Transformation}

\author{Sharon Zhang}
\orcid{0000-0002-6738-8906}
\affiliation{
 \institution{Stanford University}
 \country{USA}
 }
\email{szhang25@stanford.edu}

\author{Jiaju Ma}
\orcid{0000-0003-2880-8506}
\affiliation{
 \institution{Stanford University}
 \country{USA}
 }
\email{jiajuma@stanford.edu}

\author{Jiajun Wu}
\orcid{0000-0002-4176-343X}
\affiliation{
 \institution{Stanford University}
 \country{USA}
 }
\email{jiajunwu@cs.stanford.edu}

\author{Daniel Ritchie}
\orcid{0000-0002-8253-0069}
\affiliation{
 \institution{Brown University}
 \country{USA}
 }
\email{daniel_ritchie@brown.edu}

\author{Maneesh Agrawala}
\orcid{0000-0002-8996-7327}
\affiliation{
 \institution{Stanford University and Roblox}
 \country{USA}
 }
\email{maneesh@cs.stanford.edu}



\begin{abstract}



Motion graphics videos are widely used in Web design, digital
advertising, animated logos and film title sequences, to capture a
viewer's attention. 
But editing such video is challenging because the video provides a
low-level sequence of pixels and frames rather than higher-level
structure such as the objects in the video with their corresponding
motions and occlusions.
We present a {\em motion vectorization} pipeline for converting motion
graphics video into an SVG motion program that provides such
structure.
The resulting SVG program can be rendered using any SVG renderer
(e.g. most Web browsers) and edited using any SVG editor.
We also introduce a {\em program transformation} API
that facilitates editing of a SVG motion program to create variations
that adjust the timing, motions and/or appearances of objects.
We show how the API can be used to create a variety of effects
including retiming object motion to match a music beat, adding motion
textures to objects, and collision preserving appearance changes.

%


\end{abstract}

%
%


%
%


\begin{teaserfigure}
    \centering
    \includegraphics[width=\textwidth]{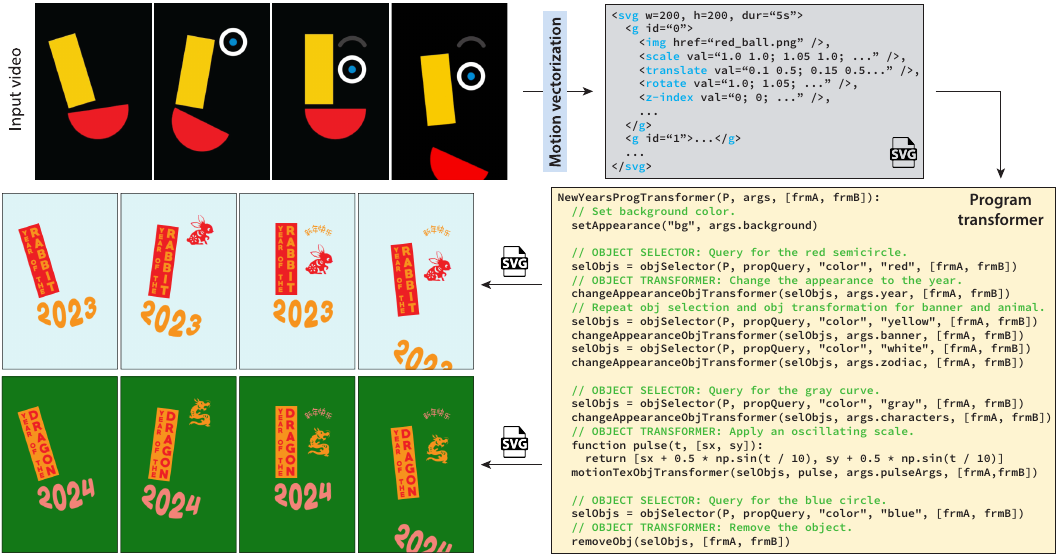}
    \caption{To edit an input motion graphics video (top left) we
      provide a pair of tools. Our {\em motion vectorization} pipeline
      converts the video into an SVG motion program that represents
      objects, their per-frame motions (scale, translate, rotate,
      skew) and their occlusion relationships (z-index). Our {\em
        program transformation} API enables programmatic creation of
      variations of the SVG motion program. Here the program
      transformer creates variations for the Chinese new year,
      selecting objects in the input video based on their color and then
      changing their appearance, matching the animal to the year and
      adding a pulsing motion texture to the Chinese characters above
      the animal icon. 
     }
  \label{fig:teaser}
\end{teaserfigure}

\maketitle


\section{Introduction}
\label{sec:intro}



Programs have proven to be useful in many areas of computer
graphics. The structure and repetition found naturally in our
surroundings, combined with the symbolic reasoning that humans use to
describe objects, can make programs particularly effective in
representing visual content. For instance, biologists use {\em
  L-systems} to model plant
structures~\cite{prusinkiewicz_lindenmayer_1996}; digital artists use
{\em shader graphs} to generate materials and
textures~\cite{Cook1984}; data analysts use {\em grammar-based APIs}
to create visualizations~\cite{2017-vega-lite,ggplot2}; and SVG is a
widely adopted {\em declarative program} format for vector
graphics~\cite{svg2}.

There are several benefits to representing visual content with a
program rather than working directly in the output space of pixels
and frames.  For one, programming languages often provide
meaningful abstractions and concepts (i.e., language primitives) that
operate at a higher level than pixels and align better with
the ways that humans think about the underlying content.
With SVG, for example, we can describe an animation as a collection of
object primitives moving in time, instead of specifying individual pixel colors over time (Figure~\ref{fig:teaser}). 
Another benefit is that programs provide meaningful control parameters.
%
SVG programs can describe the motions of objects using a sequence of
affine transforms and editing the small set of transform
parameters can generate a wide range of motions.


In this work we focus on a particular domain of visual content--namely, {\em motion graphics}--which are essentially animated graphic
designs usually consisting of shapes and typography in
choreographed motions.
Such motion graphics are ubiquitous in Web design, digital advertising,
animated logos, and film title sequences.
%
Yet, creating effective motion graphics requires expertise in crafting
eye-catching motions
and skill with animation software.
Moreover, once they have been rendered as video---the most common
format for motion graphics on the Web---they become very difficult to
edit.
Creating variations of a motion graphics video (e.g., swapping out
objects, changing the text, or retiming motions of
individual objects to music) is impractical without access to a
higher level representation.

We present tools for editing a motion graphics video by first
converting it into an SVG motion program.
Our {\em motion vectorization} pipeline identifies objects, tracks their
motions and occlusion relationships across the video, and generates an SVG motion program (Figure~\ref{fig:teaser} top row).
Our approach adapts the differentiable image compositing optimization
method of Reddy et al.~\shortcite{Reddy2020} to our tracking problem.
%
The resulting motion program can be rendered using an SVG renderer
(e.g., most Web browsers) and edited using an SVG animation editor.
To take further advantage of our representation, we introduce a {\em
  program transformation} API that allows users to programmatically
create variations of the SVG motion program.
Our approach is to treat the SVG motion program as a scene graph
composed of objects and their motions.  
%
We demonstrate how our API can be used to create a variety of effects,
including retiming object motion to match music beats, adding motion
textures (e.g., pulsing, wobbling) to objects and programmatically
changing the appearance of objects
(Figure~\ref{fig:teaser} middle, bottom rows).

\vspace{0.5em}
\noindent
In summary, we make two main contributions: 

\begin{enumerate}
\item A {\em motion vectorization} pipeline that converts a motion graphics video into an SVG motion program.
\item A {\em program transformation} API for programmatically editing SVG motion programs to create variations.
\end{enumerate}




\section{Related Work}
\label{sec:related}

\vspace{0.5em}
\noindent
{\bf \em Recovering programs from visuals.}
Because programs are such a useful representation for visual data,
graphics and vision researchers have investigated how to automatically
infer such programs from raw visual data.  This problem has been explored
in multiple visual domains, including 3D shape
modeling~\cite{Jones2020,Jones2021,Jones2022a,du2018inversecsg,Xu2021,DeepCAD,Fusion360Gallery,tian2018learning,deng2022unsupervised,Li:2020:Sketch2CAD,Li:2022:Free2CAD,kania2020ucsgnet,ren2021csg,Yu_2022_CVPR},
2D shape and layout
modeling~\cite{CSGNet,Reddy2021,Ganin2021ComputerAidedDA,seff2021vitruvion,xu2022skexgen,Ellis2018,ganin2018synthesizing},
material and texture
modeling~\cite{hu2019inverse,Hu2022a,Tchapmi:2022:Procedural,guerrero2022matformer}, extracting human motion primitives from video~\cite{Kulal2021, Kulal2022} and deconstructing visualizations~\cite{Savva2011,Harper2018,Harper2014,Poco2017}.
Deep learning is a popular technique, 
either to detect primitives
which are then combined into programs using an
optimization
process~\cite{Ellis2018,Guo6186}, to guide a search
algorithm~\cite{PlanIT,Ellis2021} or to
predict 
higher-level functions that make
programs more compact and easier to edit~\cite{Jones2021,Ellis2021}.
In our work, we leverage the visual regularity of motion graphics
videos to perform per-frame primitive detection without heavyweight
neural network machinery; we then turn these per-frame primitives into
a temporally-consistent SVG motion program via optimization.
\vspace{0.5em}
\noindent
{\bf \em Motion tracking.}  Multi-object motion tracking for
natural video is a well-studied problem~\cite{LUO2021103448,
  CIAPARRONE202061}. Many of these systems output coarse-level motion
information such as per-frame object bounding
boxes; they cannot reconstruct an input video.
Moreover, motion graphics videos tend to be relatively textureless and may contain objects that 
undergo large motions between frames. 
As a result, feature-based tracking methods such as
SIFT~\cite{Lowe2004} and KLT~\cite{LucasKanade, tomasi1991detection} are less reliable.
Recent neural network models for optical
flow~\cite{Teed,flownet} also take advantage of the
high-frequency textured nature of realistic video and are less
effective on motion graphics. In our work, we instead use neural optical flow as initialization for additional optimization or motion parameters.
%

Researchers have also developed motion tracking techniques for cartoon
style video~\cite{Sykora2009, liu2013, excol,Zhu2016}.
%
While these methods are built
for flat-colored cartoon sequences, they often produce
undesirable correspondences in motion graphics videos containing many
repeated objects (e.g., letters).  Our work is inspired by Bregler et
al.'s~\shortcite{Bregler2002}, who motion capture and retarget the exaggerated
deformations of cartoon characters.
However, they require manually annotated object contours as input,
whereas our goal is to further automate
the object detection and motion tracking process and to recover an SVG
motion program that we can retarget via a program transformation
API.




\vspace{0.5em}
\noindent
{\bf \em Layered video decomposition.}
Our work enables object-level manipulation of motion graphics video, which calls for an object-centric layered decomposition. 
Prior work in decomposing natural videos uses motion cues to generate layers based on relative depth from the camera~\cite{BrostowE99, WangAdelson1994} or on coherent camera motion~\cite{FradetPR08}. 
More recent neural methods decompose video into
layers represented as frame sequences~\cite{Lu2020,
  Lu2021} or as neural atlases~\cite{Kasten2021, Ye2022}.
Such outputs can
support appearance editing but do not enable motion
editing. Zhang et al.~\shortcite{Zhang2020} generate sprite
decompositions of cartoon videos, where each sprite is a sequence of
frames and a corresponding sequence of homographies that map between sprite and frame coordinates.
Since the appearance of each sprite can change from frame to frame, the corresponding homographies do not fully characterize the sprite motion.
They also assume a fixed depth ordering of the layers
which results in artifacts when objects change in relative depths. Our
pipeline adapts Reddy et al.'s~\shortcite{Reddy2020} differentiable compositing method 
to compute relative depth (and
motion parameters) as a function of time, allowing
for dynamic object occlusion relationships.


\section{Background}
\label{sec:background}

{\bf \em Characteristics of motion graphics video.}
Motion graphics videos are commonly composed of a set of foreground
objects, including basic shapes (e.g., rectangles, discs, etc.) and
typography moving over a static background.
The objects may occlude one another as well as split into separate
objects, or merge together into a single object.
In general, motion graphics videos may use textures and gradients to
color both the foreground objects and the background, and 
foreground objects may move and deform non-rigidly.
%
But we have found that in many contexts where motion graphics are
prevalent---e.g., Web design, animated logos,
digital advertising, film title sequences---a common stylistic choice is to use
mostly
solid-colored foreground objects undergoing affine motions over a
static background.
Sparing use of texture and photographic elements 
in combination with simpler motions can improve
legibility and make it easier to guide the viewer's gaze through the
video.
which is crucial in contexts such as
advertising.
%
We focus on converting this important class of motion
graphics video into SVG programs.






\vspace{0.5em}
\noindent
{\bf \em Structure of SVG motion programs.}
Scalable vector graphics (SVG) is a declarative programming format for
vector graphics that is widely implemented in Web browsers across a
variety of devices~\cite{svg2}.
To convert a motion graphics video into an SVG motion program we can represent
each foreground object, as an SVG group {\tt <g>} containing its
appearance {\tt <image>} and a sequence of per-frame motion
transforms.  SVG natively supports affine transforms for warping
elements with separate parameters for {\tt scale}, {\tt translate},
{\tt rotate} and {\tt skewX} and {\tt skewY}\footnote{The {\tt scale} and {\tt translate} parameters allow separate control over $x$ and $y$.}.
Each object also
includes a per-frame $z$-index depth ordering. Finally, a static
background lies at the lowest depth.
Figure~\ref{fig:teaser} shows an example of our SVG representation where we have elided
some detail to highlight the per-frame sequence of transform parameter values, ({\tt vals=...}) for one of the objects in the scene.

\section{Motion Vectorization}
\label{sec:induction}

The goal of our motion vectorization pipeline is to recover an SVG motion program
from an input motion graphics video.
%
%
The primary challenge is to identify and track each of the objects in the input video
as they appear, move, occlude one another and disappear. 
%
We use a four stage pipeline:
(1) we segment frames into regions (e.g. potential objects), (2) we
generate candidate mappings explaining how objects might move from
frame-to-frame, (3) we select the best collection of mappings
explaining the frame-to-frame movements of the objects and finally (4)
we write an SVG motion program.
%
%
Our motion vectorization pipeline builds on Reddy et al.'s~\shortcite{Reddy2020}
differentiable compositing optimization technique. We first describe
how we adapt differentiable compositing to our problem setting in
Section~\ref{sec:diffComp}; we then present each stage of our pipeline
in Sections~\ref{sec:stage1} to \ref{sec:stage6}.

\begin{figure}
    \centering
    \includegraphics[width=\linewidth]{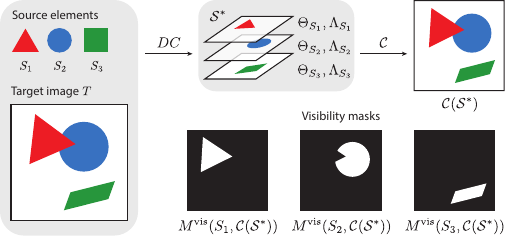}
    \vspace{-1.5em}
    \caption{Differentiable image compositing~\cite{Reddy2020}, takes a set of
      sources $\mathcal{S}=\{S_1, ..., S_N\}$ and a target image $T$ as
      input and computes a set of layering placement tuples
      $\mathcal{S}^* = \{(S_i, \Theta_{S_i}, \Lambda_{S_i})\}$ such that
      the composite image $\mathcal{C}(\mathcal{S}^*)$ matches
      $T$.
      $M^\text{vis}(S_i,\mathcal{C}(\mathcal{S}^*))$ is a binary mask
      of the visible pixels of $S_i$ after compositing.
      We extend Reddy et al.'s technique to generate affine
      transforms $\Theta_{S_i}$ rather than similarity transforms.
    }
    \vspace{-1em}
    \label{fig:diffCompFig}
\end{figure}

\begin{figure*}[t]
    \centering
    \includegraphics[width=\textwidth]{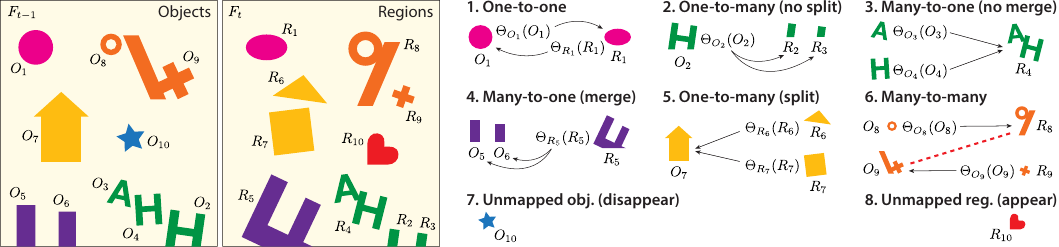}
    \caption{Eight types of mappings that can occur between objects
      $O_i$ in frame $F_{t-1}$ and regions $R_j$ in frame $F_t$.
      {\bf (1) One-to-one.} A single object $O_1$ maps to all pixels
      in a single region $R_1$ under a single affine transform
      $\Theta_{O_1}$ from the object to the region or vice versa under
      transform $\Theta_{R_1}$.  
      {\bf (2) One-to-many (no split).} A single object $O_2$ maps to
      multiple regions under a single affine transform $\Theta_{O_2}$
      from object to regions.  Since a single transform
      explains how the object moves to match all of the regions we
      consider all of them to be part of the same object (i.e.,
      the object does not split).
      {\bf (3) Many-to-one (no merge).} Two or more objects map to a
      single region, but require different affine transforms
      (e.g. $\Theta_{O_3}, \Theta_{O_4}, ...$)
      from each object to the region. Since multiple transforms are needed,
      we consider the objects as remaining distinct
      in frame $F_t$ (i.e., the objects do not merge).
      {\bf (4) Many-to-one (merge).} Two or more objects map to a
      single region under a single affine transform $\Theta_{R_5}$, from
      the region to the objects.  Since a single affine motion explains
      how the region moves to match all of the objects, we consider this a merge
      of the distinct objects.
      {\bf (5) One-to-many (split).} A single object maps two or more
      regions but require a different affine transformation to map each
      region to the object (e.g. $\Theta_{R_6}, \Theta_{R_7}, ...$). Since multiple
      transforms are required we consider the objects splitting into new distinct objects.
      {\bf (6) Many-to-many (split and merge).} Multiple objects
      map to multiple regions under differing motions. 
      Object(s) are splitting and simultaneously merging and the transforms needed to
      explain how such object(s) map to regions are ambiguous. 
      %
      {\bf (7) Unmapped object (disappear).}  When an
      object does not map to any region in the current frame
      $F_t$ we consider the object to have disappeared.
      {\bf Unmapped region (appear).} When a region does not map
      to any object in the previous frame $F_{t-1}$ we consider it a
      new object appearing for the first time. 
    }
    \vspace{-1em}
    \label{fig:mappingTypes}
\end{figure*}

\subsection{Differentiable image compositing}
\label{sec:diffComp}
Differentiable image compositing~\cite{Reddy2020} is an optimization technique originally designed to decompose a graphic pattern comprised
of discrete elements (which may partially occlude one another) into a 
layered representation (Figure~\ref{fig:diffCompFig}).
It takes in a {\em target} pattern image $T$ and a set of {\em source}
element images $\mathcal{S} = \{S_1, ..., S_N\}$
that appear in the pattern and optimizes a similarity
transform (translation, rotation, and uniform scale) for each
source element. It also computes a depth ordering so that when
the transformed elements are rendered in back-to-front order they
reproduce the target pattern.  That is,
\begin{equation}\label{eq:DC}
  \text{DC}(\mathcal{S}, T) = \{(S_i, \Theta_{S_i}, \Lambda_{S_i}) | S_i \in \mathcal{S} \},
\end{equation}
where $\Theta_{S_i}$ is the transform that places $S_i$ in
$T$, and $\Lambda_{S_i}$ is the layer z-ordering for $S_i$ in $T$ with respect to the other
elements in $\mathcal{S}$ after transforming by their
$\Theta$'s.
We refer to the resulting set of layering placement tuples as
$\mathcal{S}^* = \{(S_i, \Theta_{S_i}, \Lambda_{S_i})\}^N_{i=1}$.

With this information, we can define two additional image operators:
%
%
(1) a {\em compositing operator} $\mathcal{C}(\mathcal{S}^*)$
composites all of the transformed source elements $\Theta_{S_i}(S_i)$ in
back-to-front order according to their $\Lambda$'s;
%
%
(2) a {\em visibility mask operator} $M^\text{vis}(S_i, I)$ produces a binary
mask of the pixels of image $I$ where $S_i$ is visible. 
Importantly, $M^{\text{vis}}$ always operates in the frame space represented by $I$.
For example,
$M^\text{vis}(S_i, \mathcal{C}(\mathcal{S}^*))$ is the set of pixels of the transformed $\Theta_{S_i}(S_i)$ that are visible in $\mathcal{C}(\mathcal{S}^*)$. 
See Figure~\ref{fig:diffCompFig} for examples of both of these operators.

%
%

To apply differentiable compositing to the context of tracking objects
in motion graphics video, we have extended the optimization to 
compute an affine transformation $\Theta_{S_i}$ (translation,
rotation, non-uniform scale and skew) rather than a
similarity transform. Specifically, we add
{\tt scaleX}, {\tt scaleY}, {\tt skewX}, and {\tt skewY} as independent parameters in
the optimization.

\subsection{Stage 1: Region extraction}
\label{sec:stage1}
The first stage of our vectorization pipeline is to segment each input
frame $F_t$ into regions.  Since we focus on motion graphics with
mostly solid colored objects, as a default
we use color clustering in LAB colorspace and mark the pixels in the
cluster of the mode color as background.
Alternatively users can specify a background image if the video has a photograph, texture or colored gradient as background.
To separate the remaining foreground pixels into regions, as a default we construct
an edge map for the frame~\cite{Canny1986}
%
and then apply Zhang~\etal's~\shortcite{Zhang2009} trapped-ball
segmentation. This gives us a set of regions
$\mathcal{R}_t = \{ R_1, \ldots, R_N \}$ for each frame $F_t$.
If the foreground is textured, users can choose to skip edge detection and apply connected-components 
segementation on the foreground pixels to form regions. 
Finally, we let users manually specify pixel-level region boundaries if necessary, as noted in Section~\ref{sec:eval}.


\subsection{Stage 2: Generate candidate mapping types}
\label{sec: stage 2}
Given a set of regions for every input frame, our goal is to identify
unique foreground objects and track them between frames.  We
initialize this process at the first frame $F_1$ by treating each
region $R_i \in \mathcal{R}_1$ as an object $O_i$ so that
$\mathcal{O}_1 := \mathcal{R}_1$.
%
For each subsequent frame $F_t$, our task is to determine how objects
in the previous frame {\em map} to regions in the current frame
$\mathcal{R}_{t}$ under affine transformations.
Figure~\ref{fig:mappingTypes} shows the eight types of mappings that can occur
between objects and regions.

To determine which of these mapping types best matches 
objects in $F_{t-1}$ with regions in $F_t$, we construct an initial set of
the likeliest mapping types in the form of two bipartite graphs: (1)
the forward candidate mapping graph $\mathcal{B}_\text{fwd}$ holds
likely mappings taking objects to regions; (2) the backward candidate
mapping graph $\mathcal{B}_\text{bwd}$ holds likely mappings taking
regions to objects.
We first describe how we build the graphs and then explain how they
encode likely mappings.

\begin{figure}[t]
    \centering
    \includegraphics[width=\columnwidth]{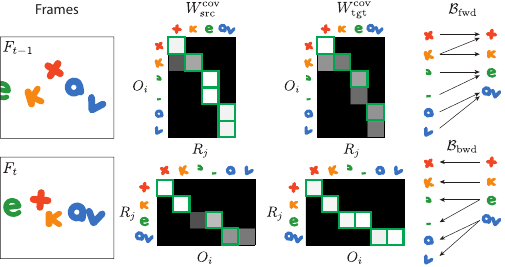}
    \vspace{-2em}
    \caption{To build the forward candidate mapping graph
      $\mathcal{B}_\text{fwd}$ (top row), we consider each edge $(O_i,
    R_j)$ from object $O_i$ to region $R_j$ and compute coverage
    weights $W^\text{cov}_\text{src}(O_i, R_j)$ and
    $W^\text{cov}_\text{tgt}(O_i, R_j)$. We retain only highest non-zero
    weighted edges in the graph for each object -- highlighted in green in the matrices, one per row.
    We similarly build the backward mapping graph
    $\mathcal{B}_\text{bwd}$ (bottom row), but flip the direction of the
      edges $(R_j, O_i)$ to run from region $R_j$ to object $O_i$ with
      the coverage weights similarly inverted
      $W^\text{cov}_\text{src}(R_j, O_i)$ and
      $W^\text{cov}_\text{tgt}(R_j, O_i)$ (bottom row).
      %
          }
    \label{fig:buildFwdBwdGraphs}
\end{figure}

\begin{figure}[t]
    \centering
    \includegraphics[width=\linewidth]{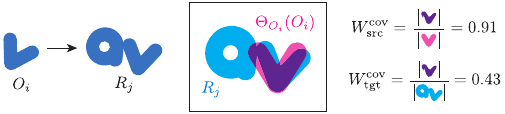}
    \caption{For edge $(O_i, R_j)$, we compute coverage weights $W^\text{cov}_\text{src}$ and
      $W^\text{cov}_\text{tgt}$ by first transforming the source
      object $O_i$ to form $\Theta_{O_i}(O_i)$.
      $W^\text{cov}_\text{src}$ is the area of the visible overlap between $\Theta_{O_i}(O_i)$ and $R_j$ (purple)
      as a percentage of the visible area of the transformed object $\Theta_{O_i}(O_i)$ (pink or purple).
      $W^\text{cov}_\text{tgt}$ is the area of the overlap (purple)
      as a percentage of the visible area of the target region $R_j$ (cyan or purple).
    }
    \vspace{-1em}
    \label{fig:cov_score}
\end{figure}

\vspace{0.5em}
\noindent
{\bf \em Build candidate mapping graphs.}
Figures~\ref{fig:buildFwdBwdGraphs} and~\ref{fig:cov_score} show how we build $\mathcal{B}_\text{fwd}$ and $\mathcal{B}_\text{bwd}$.
For $\mathcal{B}_\text{fwd}$, we first apply differentiable
compositing as $\text{DC}(\mathcal{O}_{t-1}, F_t)=\mathcal{O}^*$,
treating $\mathcal{O}_{t-1}$ as the set of source elements and the
current frame $F_t$ including all of its regions $\mathcal{R}_t$,
as the target image.
Then, for each object $O_i \in \mathcal{O}_{t-1}$, we consider each
region $R_j \in \mathcal{R}_t$ and compute a {\em source} coverage
weight as
\begin{equation}
  W^\text{cov}_\text{src}(O_i, R_j) = \frac{|M^\text{vis}(O_i, \mathcal{C}(\mathcal{O}^*)) \cap M^\text{vis}(R_j, F_t)|}{|M^\text{vis}(O_i, \mathcal{C}(\mathcal{O}^*))|}.
  \label{eq:srcWeight}
\end{equation}
%
%
This weight measures the visible overlap between the transformed object and the 
region as a percentage of the visible area of the transformed object (Figure~\ref{fig:cov_score}). 
%
We add the highest non-zero weighted edge $(O_i, R_j)$ to the forward
graph $\mathcal{B}_\text{fwd}$ (top left weight matrix in
Figure~\ref{fig:buildFwdBwdGraphs}).
%
%
Similarly, for each region $R_j$, we consider each object
$O_i$
and compute a {\em target} coverage weight as
\begin{equation}
  W^\text{cov}_\text{tgt}(O_i, R_j) = \frac{|M^\text{vis}(O_i, \mathcal{C}(\mathcal{O}^*)) \cap M^\text{vis}(R_j, F_t)|}{|M^\text{vis}(R_j,F_t)|}.
    \label{eq:tgtWeight}
\end{equation}
%
This weight measures the visible overlap between the transformed object and the 
region as a percentage of the visible area of the region (Figure~\ref{fig:cov_score}).
We add the highest non-zero weighted edge $(O_i, R_j)$ to
$\mathcal{B}_\text{fwd}$ if it has not already been added to the graph
(top right weight matrix,
Figure~\ref{fig:buildFwdBwdGraphs}).

The backward graph is built in exactly the same way except that we
treat the regions $\mathcal{R}_t$ as source elements and the previous
frame $F_{t-1}$ as the target in the differentiable compositing
optimization to compute $\text{DC}(\mathcal{R}_t, F_{t-1}) =
\mathcal{R}^*$.  For the coverage weights computations
(Equations~\ref{eq:srcWeight} and~\ref{eq:tgtWeight}), we similarly
flip the computation treating regions $R_j$ as sources and objects
$O_i$ as targets and replace $F_t$ with $F_{t-1}$ (bottom row,
Figure~\ref{fig:buildFwdBwdGraphs}).

In practice, we have found that DC is sensitive to
the initial placement of source elements.
Therefore, we initialize the source placement using shape context~\cite{Belongie2006}, 
optical flow~\cite{Teed} and RANSAC to estimate how each object
(or region) moves to $F_t$ (or $F_{t-1}$).
%
Note also that when we use DC, we save the
resulting sets of layering placement tuples $\mathcal{O}^*$ and
$\mathcal{R}^*$ for use in later stages of our pipeline.


\vspace{0.5em}
\noindent
{\bf \em  Extract candidate mappings.}
%
%
%
The forward and backward candidate mapping graphs encode multiple
candidate mappings.  To extract the individual candidate mappings
from either of these graphs, we first consider each connected
component of the graph.  We treat any such component that is
one-to-one, one-to-many, or many-to-one (i.e., the component
contains exactly one object or exactly one region) as a candidate
mapping.
If the component 
\setlength{\columnsep}{6pt}
\begin{wrapfigure}[8]{r}{1.30in}
      \centering 
      \vspace{-1.25em}
      \includegraphics[width=1.3in]{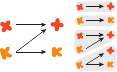}
      \vspace{-2.25em}
      \caption{Breaking a many-to-many component of $\mathcal{B}_\text{fwd}$.
        }
  \label{fig:many-to-many}
\end{wrapfigure}
forms a many-to-many graph, we further break it into 
pieces (see inset) as follows. For each
node (object or region) in the component, we form a subgraph that
includes all edges the node is part of.  Each resulting
subgraph is then either a one-to-one, one-to-many, or many-to-one mapping
candidate.

As shown in Figure~\ref{fig:mappingTypes}, many-to-many mappings
are ambiguous because they require object(s) to simultaneously split
and merge. In practice, we have found that such split-merges
are rare for the kind of motion graphics videos we focus on
in this work.
Thus, our approach is to force our algorithm to explain many-to-many
mappings as a combination of one-to-one, one-to-many, or many-to-one
mappings.
Figure~\ref{fig:selection} shows the complete set of mappings we extract from $\mathcal{B}_\text{fwd}$
and $\mathcal{B}_\text{bwd}$ for the example in
Figure~\ref{fig:buildFwdBwdGraphs}.

\subsection{Stage 3: Select best collection of mappings}
\label{sec:stage3}

\begin{figure}
     \centering
     \includegraphics[width=\columnwidth]{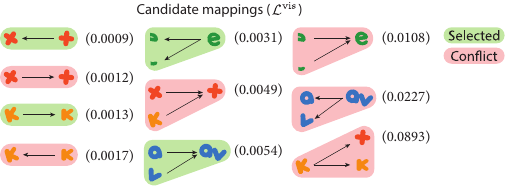}
     \vspace{-1.5em}
     \caption{We compute penalty scores $\mathcal{L}^\text{vis}$ for each candidate mapping and then select the best conflict-free set  of mappings using a greedy approach. 
     }
     \label{fig:selection}
     \vspace{-1em}
 \end{figure}

To select a set of mappings that best explain how objects move from frame $F_{t-1}$ to $F_t$ 
we first score each candidate mapping we obtain in stage 2 using a
visibility-based penalty loss.
Suppose $H$ is a candidate mapping type extracted from the forward
graph, and $\mathcal{O}_{t-1}^H$ and $\mathcal{R}^H_{t}$ are the set of
object(s) and region(s) in $H$. We define the visibility loss
$\mathcal{L}^\text{vis}$ as a masked $L_2$-norm of color differences
between the composite image $\mathcal{C}(\mathcal{O}^*)$ of the
transformed
and layered objects,
and the current frame $F_t$. That is,
\begin{equation}
  \mathcal{L}^\text{vis}(\mathcal{O}^H_{t - 1}, \mathcal{R}^H_t) = || (\mathcal{C}(\mathcal{O}^*) - F_t) \otimes M^\text{all}  ||_2,
  \label{eq:lvis}
\end{equation}
where $\otimes$ denotes pixel-wise multiplication and $M^\text{all}$ is a mask
\begin{equation}
  M^\text{all} = \left( \bigcup_{O_i \in \mathcal{O}^H_{t - 1}} M^\text{vis}(O_i, \mathcal{C}(\mathcal{O}^*)) \right) \cup \left( \bigcup_{R_j \in \mathcal{R}^H_t} M^\text{vis}(R_j, F_t) \right),
  \label{eq:mall}
\end{equation}
consisting of 
the union of the visible pixels of all of the transformed
objects $O_i \in \mathcal{O}^H_{t - 1}$ (first term) with the union of all of the
regions $R_j \in \mathcal{R}^H_t$ (second term).
This loss is minimized when the pixels of the transformed objects in
$H$ match those of corresponding regions in $H$ and there are no
mismatched pixels.
Similarly, if $H$ is a candidate mapping type from the backward graph,
we compute the penalty score as
$\mathcal{L}^\text{vis}(\mathcal{R}^H_t, \mathcal{O}^H_{t-1})$, while
replacing $\mathcal{O}^*$ with $\mathcal{R}^*$ and $F_t$ with
$F_{t-1}$ in Equations~\ref{eq:lvis} and~\ref{eq:mall}.
%
In particular, the visibility loss differs from the coverage weights (Section~\ref{sec: stage 2}) as it evaluates the {\em color} appearance of an entire mapping rather than object-region alignment.
Figure~\ref{fig:selection} shows the penalty scores for the mappings
we extracted for the example in
Figure~\ref{fig:buildFwdBwdGraphs}.

We next select a set of conflict-free mappings from our set of
candidates that collectively best explain how objects move, appear, or
disappear between frames $F_{t-1}$ and $F_t$.  A pair of candidate mappings
are in conflict if they include the same
object or region (Figure~\ref{fig:selection}).
Starting with the complete set of candidate mappings, we repeatedly select the candidate with the lowest penalty score and remove all conflicting candidates from the set.
We stop when the candidate mapping set is empty, or the lowest score of the
remaining candidates is greater than a threshold $\epsilon$.
We have found that
$\epsilon=0.1$ gives good results across all our examples.

\begin{figure}
    \centering
    \includegraphics[width=\linewidth]{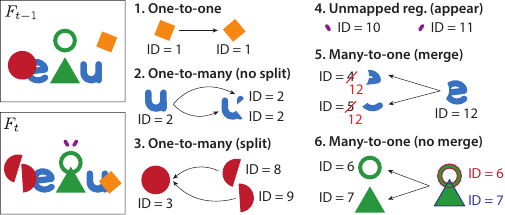}
    \vspace{-1.5em}
    \caption{Propagating IDs based on mapping type. For one-to-one and
      one-to-many (no split) mappings, we assign all pixels of the
      corresponding region(s) the ID of the object.  For one-to-many
      (split) and unmapped region (appear) mappings, we create new IDs
      and label the pixels of each region with a different ID.  For many-to-one (merge) mappings, we create a new ID to assign to
the pixels of the region and then relabel all previous instances of
the corresponding objects in the mapping to this new ID.
For many-to-one (no merge) mappings, we assign the IDs of
each object $O_i$ in the mapping to the corresponding pixels in
$\Theta_{O_i}(O_i)$.
    }
   \vspace{-1em}
    \label{fig:IDprop}
\end{figure}

Finally, we propagate object IDs from the previous frame objects
$\mathcal{O}_{t-1}$ to current frame regions $\mathcal{R}_t$ based on
the selected mappings as shown in Figure~\ref{fig:IDprop}. Anytime an object disappears we do not propagate its ID to any subsequent regions. Thus, objects which become completely occluded will re-appear with a new ID by default, though this can be easily changed with user input (Section~\ref{sec:eval}).  During
this process we also keep track of a \textit{canonical image}
for each object. When an object first appears, we save its labeled
pixels as its canonical image. Every time an object appears unoccluded
and covers a larger region of pixels in a subsequent frame, we update
that canonical appearance by replacing the entire canonical image. Thus we maintain a high-resolution appearance for each object.



\subsection{Stage 4: Write an SVG motion program}
\label{sec:stage6}

In the final stage, we refactor the frame-to-frame affine
motion transforms for each object into an affine transform mapping the
object's canonical image to each frame.
This motion refactorization could be obtained by multiplying 
the frame-to-frame transforms or their inverses.
In practice, we have found that we can further increase motion
accuracy by re-running the DC optimization
using the canonical images as the source and the
corresponding labeled pixels in each frame as the
target.
%
%
Finally, we write out a SVG motion program with a static
background image and a set of foreground objects, each represented by
a canonical image, a per-frame sequence of affine transforms placing
the canonical image in the frame, and a per-frame z-index depth for
the object.



\begin{figure*}[t!]
    \centering
    \includegraphics[width=\textwidth]{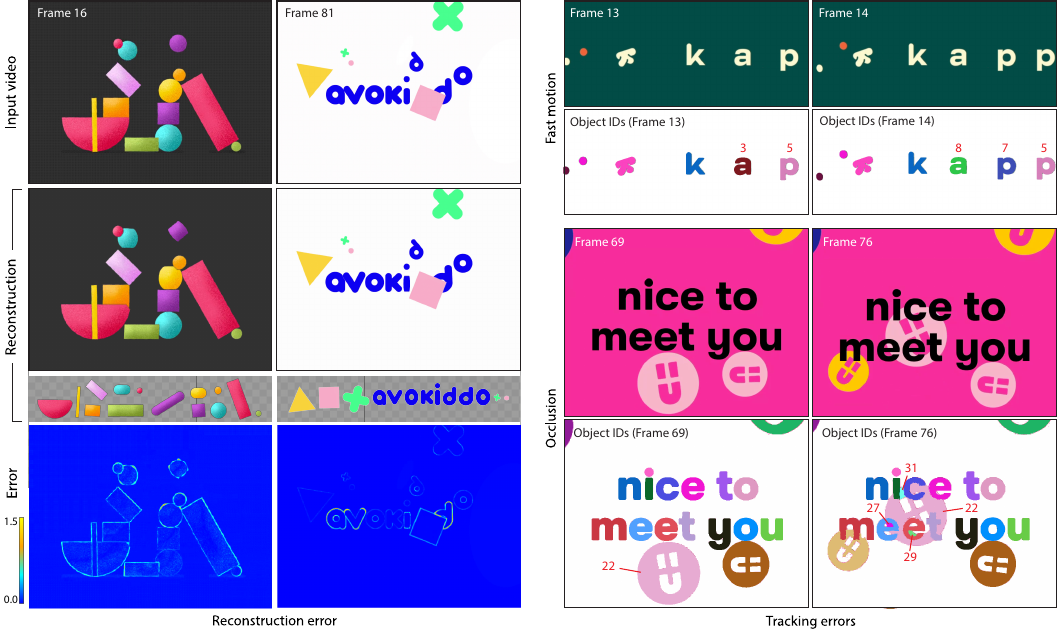}
        \vspace{-1.75em}
        \caption{{\bf Left:} Reconstruction errors ($L_2$ RGB difference)
          between frames of input motion graphics videos ({\em 5k,
            avokiddo}) and the corresponding frames rendered with the
          SVG motion program generated by our vectorization
          pipeline. {\bf Right:} Tracking errors due to fast object motions
          and occlusions.  {\bf Right Top:} The {\em kapptivate} input video
          contains characters translating quickly right to right. In
          frame 13 the `a' is correctly assigned object ID 3, but in
          frame 14 it is incorrectly assigned a new object ID 8.  This
          occurs because the leftmost `p' in frame 14 is the closest
          similar looking region to the `a' in frame 13 but the
          candidate mapping between the `a' and the `p' is rejected as
          being too low quality.  The `p' in frame 13 is also
          incorrectly mapped to the rightmost `p' in frame 14 for
          similar reasons, while the leftmost `p' in frame 14 is
          incorrectly assigned a new object ID 7 since it remains
          unmatched. Thus this example yields 2 one-to-one mapping
          errors and 1 unmatched region (appear) error. {\bf Right Bottom:}
          In the {\em lucy} video object 22 is correctly
          tracked before frame 76 (we visualize it in frame 69 to show
          the complete unoccluded object).  In frame 76 occlusions
          alter the visibility of the corresponding region so much
          that a one-to-many (no split) mapping is misidentified as a
          one-to-many (split) mapping and the additional regions are
          given brand new IDs 27, 29 and 31.
        }
        \vspace{-1em}
    \label{fig:eval}
\end{figure*}

\section{Results: Motion Vectorization}
\label{sec:eval}

Figure~\ref{fig:teaser} shows an abstracted example of the SVG motion
program our vectorization pipeline recovers from an input motion graphics
video. 
We apply our motion vectorization pipeline on a test set of 38 motion graphics
videos sourced from the Web,
with many containing occlusions or fast object motion.  A few 
videos include textures, photographic elements or color gradients in the foreground or
background. 
Table~\ref{tab:tracking_eval} (Appendix~\ref{sec:appendix}) gives
more detail about these videos and the supplemental website provides
complete running SVG motion programs 
for all of them.

We first consider the reconstruction error between frames of the input
motion graphics videos and corresponding frames produced by the SVG
motion programs.  Overall, the average $L_2$ RGB error across our 
test set is $0.0086$.
Slight reconstruction errors appear mostly at edges of objects due to
small inaccuracies in transform parameters, noise, compression or
anti-aliasing (Figure~\ref{fig:eval} left). 
As a comparison we also use the sprite-from-sprite decomposition method~\cite{Zhang2020}. Sprite-from-sprite successfully decomposes the 30 test videos and runs out of memory on the rest. The average $L_2$ RGB reconstruction error for sprite-from-sprite on this subset of videos is 0.018, compared to 0.0079 using our method.
See supplemental materials A for a more detailed discussion of this comparison.

We also compute the number of tracking errors in each video.
We define a tracking error as any time a mapping from objects in frame
$F_{t-1}$ to regions in $F_t$ is incorrect with respect to a manually
annotated set of ground truth mappings.
%
Table~\ref{tab:tracking_eval} (Appendix~\ref{sec:appendix})
shows the total number of such tracking errors as well as the count of
errors amongst each mapping type for all the videos in our test set.

We find that 24 videos in our test set contain no tracking errors at
all, even as some of them contain fast motion, occlusions, or both.  The
remaining 14 videos all contain 15 errors or fewer.
%
Across
all the videos, 75\% of the tracking errors occur in one-to-one
mappings.  Such errors are often due to fast motion and occlusions
when objects enter or exit the frame (Figure~\ref{fig:eval} right
top).
%
%
The next most
common tracking error type, at 21\%, is incorrect one-to-many
(no-split) mappings. Such errors often occur when objects occlude one
another and the mapping is misidentified as a one-to-many (split)
(Figure~\ref{fig:eval} right bottom).
%
Two of the three remaining tracking errors occur when many-to-one (no merge)
mappings are misidentified as many-to-one (merge) mappings. In these
cases the video contains similarly
colored overlapping objects that move in unison, so our pipeline merges them into one
object. The final tracking error occurs when a newly appearing
region is incorrectly mapped to an existing object.  The unmatched
region (appear) mapping is misidentified as a one-to-one mapping (example in 
Figure~\ref{fig:eval} top right).
Our test set did not produce errors of the other four mapping types.

%

\vspace{0.3em}
\noindent
{\bf \em Correcting tracking errors.}  Most tracking errors
    are easily fixed by reassigning object IDs to regions.  For
    instance if a region was assigned object ID 3 but should have
    been assigned object ID 7, we can manually relabel it.  We provide
    a programmatic interface for such reassignment.  An error in a
    many-to-one (no merge) mapping can require breaking the pixel mask
    of a region into multiple regions.  In this case users can
    manually specify the pixel boundaries of each region in the frame
    where the error appears in Stage 1 of our pipeline to enforce the
    correct region boundaries.  We found this correction to only be
    necessary for two videos ({\em shapeman}, {\em confetti}) in our test
    set. In general however, because our pipeline produces relatively
    few tracking errors they can often be corrected very
    quickly. 

\vspace{0.3em}
\noindent
    {\bf \em Discussion.} 
The SVG motion programs produced by our vectorization pipeline provide a
representation of motion graphics videos that can be rendered
using a SVG renderer, including most Web browsers.
In addition, the
motion programs can be edited using a SVG animation editor. We have
built SVG motion program importers for Adobe After
Effects~\cite{christiansen2013adobe} and
Blender~\cite{blender3d}. Such editors allow users to manually
customize the motion and appearance of the objects using a graphical
interface they may already be familiar with (see supplementary video).

\section{Motion Program Transformation}
\label{sec:tranformation}



Our {\em program transformation} API lets users programmatically
express different ways of manipulating an SVG motion program to
generate variations of it.
%
%
Our approach is to treat the SVG motion program as a scene graph
that describes the motions of objects over time.
Our API adopts a well known-design pattern for working with a scene graph
via two types of methods;
(1) {\bf \em state queries} that look up information
about the objects and events in the scene, and (2) {\bf \em operators}
that modify the appearance or motion of objects.
A transformation program typically starts by querying for a set of
objects based on their properties (e.g. red colored objects) or the
events they participate in (e.g. collisions) and then applies one or
more operators to modify the selected objects.
This design pattern of querying and then modifying a scene graph is
often used in game engines (e.g., Unity~\cite{unity2023manual})
as well as Web APIs (e.g. jQuery~\cite{jquery2023doc},
D3~\cite{bostock2011d3}, CSS~\cite{css2023doc} and
Chickenfoot~\cite{Bolin2005}) that treat the DOM as a scene graph.

We describe the methods of our program transformation API
(Sections~\ref{sec:queries} and~\ref{sec:operators}) and briefly describe how
we can use them to build a variety of higher-level transformation effects
(Section~\ref{sec:effects}).
The supplemental materials B provides additional details about our API as well as multiple code
examples.
%
While our proof-of-concept implementation of the API enables all of
the examples that follow, it is meant to minimally demonstrate our approach.
In practice, it could be extended to include additional state queries and
operators as necessary.

\begin{figure*}
  \centering
      \vspace{-1em}
    \includegraphics[width=\textwidth]{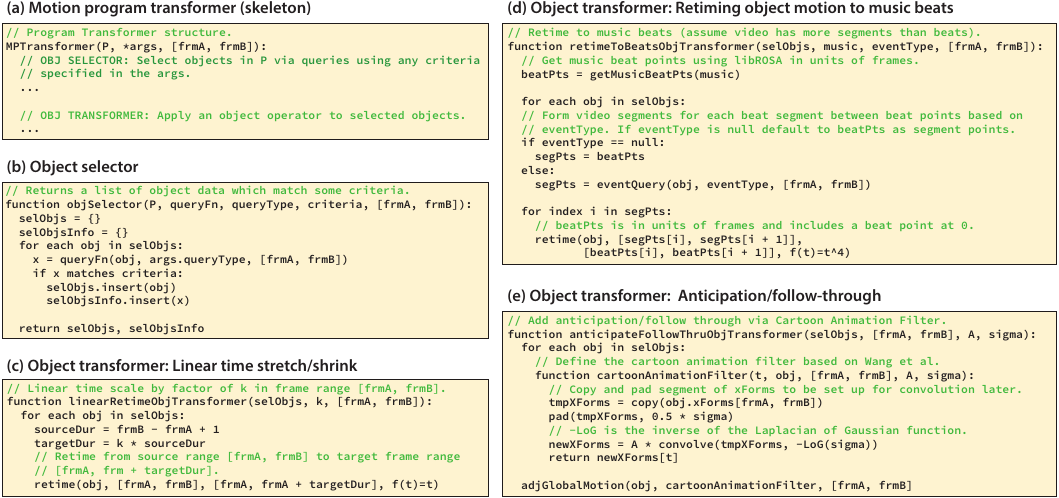}
    \caption{The general structure of motion program transformer (a)
      takes an SVG motion program {\tt P} as input and alternates
      object selector blocks with object transformer blocks to modify
      the SVG program. The object selector function {\tt objSelector} (b)
      selects one or more objects for transformation. It first runs
            {\tt queryFn} (i.e., either {\tt propQuery} or {\tt eventQuery}) using the specified {\tt
              queryType} (i.e., {\tt color}, {\tt collisionFrames}) and then filters the objects to only those 
            that match the specified {\tt criteria}. The object transformers adjust the timing (c, d) motion (e) or appearance of a set of selected objects {\tt selObjs}. See the supplemental material B for additional examples of object transformers we have built to achieve a variety of effects.}
    \vspace{-0.5em}
    \label{fig:code}
\end{figure*}







\subsection{Program Transformation API: {\em State Queries}}
\label{sec:queries}

State queries retrieve properties or events for a specific object,
over a range of frames:

{\small 
\begin{description}
\item[\texttt{propQuery(obj, propType, [frmA, frmB])}:] Returns a property of \texttt{obj} for each frame
  in {\tt[frmA, frmB]} based on
  \texttt{propType}. Property types include: {\tt all}, {\tt color}, {\tt position}, {\tt size},
         {\tt velocity}, etc.
\end{description}
}

{\small 
\begin{description}
\item[\texttt{eventQuery(obj, eventType, [frmA, frmB])}:] Returns a
  list of events \texttt{obj} is involved in over the range of frames
  {\tt[frmA, frmB]} based on \texttt{eventType}.  Event types include:
  \texttt{heldFrames}, \texttt{collisionFrames}, \texttt{motionCycleFrames}, etc.
\end{description}
}

To handle property queries, our API internally computes the chosen
property for the object from our motion program representation. 
For example, to compute the {\tt color} property of an object it clusters the pixels of the canonical
image in color space and returns the color of the largest cluster for
each frame in the frame range.
Properties that vary based on the motion (e.g., {\tt position}, {\tt
  size}, {\tt velocity}) are computed using the objects motion
transform and reported in the global coordinates of the video frame.
The {\tt all} property type returns all objects that appear in motion program over the frame range.

To handle event queries, our API internally processes the motion of
the object to find frames when the chosen event type occurs.
For example, to identify {\tt heldFrames} we look for successive
frames of the object where the motion transform from the canonical
image to the frame placement remains fixed and return a list of all
such frames.
To identify {\tt collisionFrames} we look for frames where
the closest distance between the object boundary and another object boundary is below
a threshold (e.g. the objects touch) and at least one of the objects experiences a large change in velocity.
%
%
%
The API returns a list of collisions including
the other object(s) involved and the points of contact on each object.
To identify {\tt motionCycleFrames} we look for peaks in the
autocorrelation of motion parameters (translation, rotation, scale skew)
of the object and return a list of the
corresponding frames.

\subsection{Program Transformation API: {\em Operators}}
\label{sec:operators}

Our API provides operators to modify the appearance or motion of a
specific object over a range of frames including:

{\small
\begin{description}

      \item[\texttt{retime(obj, [sFrmA, sFrmB], [tFrmA, tFrmB], easeFn[t])}:]
         Linear\-ly remap motion transforms in source frame range
         {\tt [sFrmA, sFrmB]} to target frame range {\tt [sFrmA,
             sFrmB]}. Then resample the transforms in the target frame
         range using easing function {\tt easeFn[t]}.

      \item[\texttt{adjLocalMotion(obj, xformFn[t], [frmA, frmB])}:]
        Adjust motion of {\tt obj} in local coordinate frame (i.e., of
        canonical image), over the range of frames {\tt[frmA, frmB]}
        based on affine transforms generated by linearly sampling {\tt
          xformFn[t]} in the range {\tt [0,1]}. 
        This method post-multiplies canonical-to-frame transform of {\tt obj}.


      \item[\texttt{adjGlobalMotion(obj, xformFn[t], [frmA, frmB])}:]
        Adjust motion of {\tt obj} in global coordinate frame (i.e., of
        video frame), over the range of frames {\tt[frmA, frmB]}
        based on affine transforms generated by linearly sampling {\tt
          xformFn[t]} in the range {\tt [0,1]}. This method pre-multiplies the canonical-to-frame transform of {\tt obj}.

    \item[\texttt{changeAppearance(obj, newAppearance, [frmA, frmB])}:]
    Set canonical image of \texttt{obj} to \texttt{newAppearance} for frames in {\tt[frmA, frmB]}.
\end{description}
}
In addition to the operators listed here, our API provides basic
operators for creating new objects, deleting objects, copying motions, setting the
motion transforms (rather than adjusting them via pre- or
post-multiplication), etc.  


\vspace{0.5em}
\noindent
Figure~\ref{fig:code} shows the the general pattern of a motion
program transformer, written with our API. An {\tt objSelector}
code block (or function) selects one or more objects for transformation
using a {\tt propQuery} or {\tt eventQuery}.
%
An {\tt objTransformer} code block (or function) then applies one or
more operators to change the timing, motion or appearance of the
selected object(s).
%
For example, to transform all of the red colored objects to blue, the
{\tt objSelector} function would run a {\tt propQuery} to obtain the
{\tt color} of each object and then select out the red ones. Then the
{\tt objTransformer} code block would use {\tt changeAppearance} to
set the color of the selected objects to blue.
%
%
%


\subsection{Higher-level object transformer effects}
\label{sec:effects}

Using our motion program transformation API
we have built a variety of {\tt objTransformer}
functions
that each produce a different, higher-level effect
on the timing, motion or appearance
of objects (e.g. anticipation/follow-through, motion textures).
Several of these transformers implement motion adjustments commonly found in other animation editing systems~\cite{Ma2022, Kazi2014, Kazi2016MotionAmp}.
Importantly, the functions in our API are designed to compose with one another and
facilitate the creation of many variations of a motion graphic, thereby
supporting iterative design and exploration. 
Figure~\ref{fig:code} provides code for a few object transformers, and supplemental materials B includes code for all of them.
%
The supplemental website also includes multiple example SVG motion
programs transformed by each of the higher-level effects described
here that can be executed in a Web browser. The following sections give a brief overview of the types of object transformers.

{\bf \em Retiming.} These object transformers manipulate an individual object timeline. This includes functions that linearly stretch or shrink the time scale of an object, apply slow in/out easing, retime object motions to reference audio beats, etc. See Figure~\ref{fig:code}c and~\ref{fig:code}d for examples.

{\bf \em Spatial motion adjustment.} These adjustment object transformers manipulate how an individual object moves across the frame. This includes functions that add anticipation/follow-through (Figure~\ref{fig:code}e) and functions that apply motion textures (e.g. wobbling or pulsing) to an existing motion.

\begin{figure*}
    \centering
    \includegraphics[width=\textwidth]{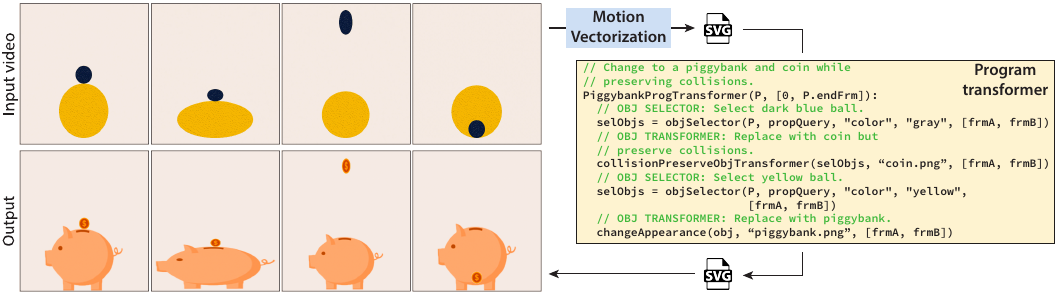}
    \caption{Changing appearance while preserving collisions. This
      input video contains two balls that interact with one another
      with the dark blue ball bouncing around outside and inside the
      yellow ball. The program transformer changes the blue ball into
      a coin that is smaller than the blue ball. It then uses the {\tt
        collisionPreserveObjTransformer} to adjust the motion of the
      smaller coin so that the collision points are maintained with
      the yellow ball. Finally it changes the appearance of the yellow
      ball to a piggy bank with the body of the bank the same size as
      the yellow ball.}
    \label{fig:collision}
\end{figure*}

{\bf \em Appearance adjustment.} The \texttt{changeAppearance} object transformer updates the appearance of a given object by replacing the canonical appearance of an object with a new image. One unintended consequence of an appearance change is that collisions between objects may be affected. For instance, naively changing the dark blue circle in Figure~\ref{fig:collision} to a smaller-sized coin would not maintain collisions between the smaller coin and the yellow circle. Since collisions are often important events in a video, we also allow for {\em collision-preserving} appearance changes. This type of appearance change uses event queries to find \texttt{collisionFrames} and then applies local motion adjustments to best preserve the original collisions at those frames. 

\section{Results: Motion Program Transformation}
\label{sec:resultsTransformation}


By combining \texttt{objSelector} and \texttt{objTransformer} blocks, we can create a variety of motion graphic variations. Figure~\ref{fig:teaser}, Figure~\ref{fig:collision} and Figures~5--6 in supplemental materials A
show examples where we have composed multiple {\tt objSelector} and
{\tt objTrans\-former} blocks to generate complex variations of retiming, spatial motion adjustment and appearances changes.
%
Executable SVG motion programs and program transformer code for other additional examples with retiming, spatial motion adjustments and appearance adjustments are provided in the supplemental website.
We encourage readers to browse the examples to see the breadth of
different transformations and variations that can be achieved with our
motion program transformation API.
%

\begin{figure}
    \centering
    \includegraphics{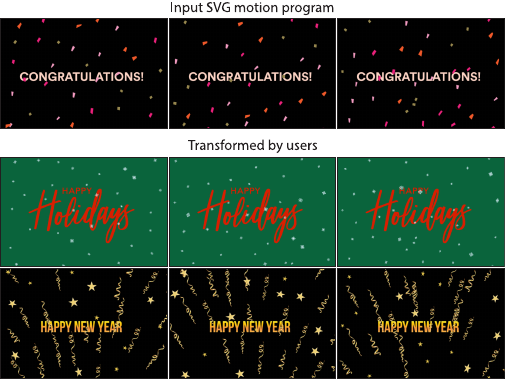}
    \vspace{-2em}
    \caption{We asked user study participants to use our transformation API to repurpose a digital card with confetti falling down (top row). 
    One participant created a happy holidays card with falling snow (middle). Another created a new years card reversing the falling motion to create streamers and stars. 
    }
        \vspace{-1em}   
    \label{fig:user_study}
\end{figure}

\vspace{0.5em}
\noindent
{\bf \em Usability evaluation.}
To further evaluate the usability of our program transformation API,
we asked 10 people (all experienced Python coders, 5 familiar with
query-then-operate design pattern) to use the API to programmatically
create a variation of an SVG motion program
(Figure~\ref{fig:user_study}). We first gave each participant a 30
minute tutorial (a combination of oral instruction and a Colab
notebook) explaining how to use the API. We then gave them 15 minutes
to write their own program transforming an animated digital card
into one suitable for a different occasion.

All participants successfully wrote a transformation program
containing two or more object queries and transformations. On a 5
point Likert scale (1 = {\em very hard}, 5 = {\em very easy}) they all
rated the query-then-operate pattern as {\em easy} or {\em very easy}
to understand.  Two participants who were familiar with the design
pattern compared the structure of our API to SQL and other scene-graph
based content creation APIs like Maya~\cite{maya} and
MotionBuilder~\cite{motionbuilder}. 
Multiple participants stated in free-response feedback that the API was
"intuitive to understand," "lightweight and natural," and "easy to
use."

Many participants liked the expressivity of the API.  Nine
participants noted that the API was flexible enough to accomplish the
edits they wanted to make.  One participant liked ``how powerful the
API is while still being easy to use," further commenting that ``it
covered a lot of possible transformations within relatively simple
operations." Another wrote that the programmatic approach of our API
``would be especially useful for mass producing animations or images
that still look customized" and ``[they] would welcome [the]
programmatic approach compared to painful and arduous manual process
of doing it through interfaces like InDesign."
%
Overall, this feedback suggests that users familiar with
programming are able to use our transformation API to easily
produce variations of a SVG motion program.

\section{Limitations and Future Work}
\label{sec:lim}

Our work enables editing of motion graphics video by first converting
the video into an SVG motion program and then using motion program
transformers programmatically create variations. However there are a
few limitations that warrant future work.

\vspace{0.5em}
\noindent
    {\bf \em Lifting assumptions on input video.}  Our work focuses on
    motion graphics video with a static background and solid-colored,
    lightly textured or gradient-filled objects undergoing affine
    motions. Extending our approach to handle natural video containing
    moving backgrounds with highly textured, photographic foreground objects
    undergoing deformable motions, may be possible using recent video
    matting techniques~\cite{Lu2021,Kasten2021}. Handling non-affine
    motions within our pipeline would require modifications to the
    differentiable compositing optimization
    (Section~\ref{sec:diffComp}) to account for the deformations.

\vspace{0.5em}
\noindent
{\bf \em Vectorizing canonical images.} Our SVG motion programs
represent the appearance of each object using a canonical
image. Converting these canonical images into a vector representation
(e.g., composed of paths, shapes, gradients, etc.) would bring the
benefits of a higher-level abstraction to the appearance of the
objects in addition to their motions. Techniques for converting images
into vector representations~\cite{orzan2008diffusion, Reddy2021} is an active
area of work that might be adapted to this context.

\vspace{0.5em}
\noindent
{\bf \em Higher-level program abstraction based on gestalt
  principles.}
Our SVG motion programs represent motion graphics video using
abstractions (e.g., objects) and controls (e.g., affine transform
parameters) that are more meaningful than pixels and frames of
video. One way to provide further meaningful abstraction might be to
group objects based on perception and gestalt principles. For example
if a motion graphic contains objects (e.g., letters) that move together
and are near one another, they might be
grouped together to form a higher-level composite object (e.g., a
word). Such higher-level grouping could further facilitate program transformation
as changes and adjustments could be applied to the composite objects.

\vspace{0.5em}
\noindent
{\bf \em GUI for motion editing.} Our system enables users to work with a
programmatic representation of motion graphics video rather than
pixels and frames. However, we have not developed a graphical user
interface for editing the resulting SVG motion programs. Indeed, we
believe many different GUIs could be built using our motion program
representation and our program transformation API.  One approach that
may be especially fruitful is to extend the bidirectional SVG editing interface of
Sketch-n-Sketch~\cite{Hempel2019}, 
so that direct manipulation changes to the graphics are immediately
reflected in the SVG representation and vice versa. Inferring how
direct, graphical manipulations should affect an underlying motion
program is an important direction for future work.



\section{Conclusion}
\label{sec:conc}

While motion graphics videos are prevalent on the Web today, they are
difficult to edit because they are simply a collection of pixels and frames.
We have presented a motion vectorization pipeline that converts such video into
a SVG motion program that represents the video as objects moving over time.
We further provide a motion program transformation API that enables
programmatic editing of the resulting SVG programs to create variations
of the timing, motions and object appearance.
We believe that these tools can allow users to more easily explore
motion graphics design options by borrowing from widely-available
motion graphics video examples and that they open the door to
dynamically adapting the graphics to the preferences of the viewer.

\begin{acks}
We thank Lvmin Zhang for valuable discussions on sprite-from-sprite. We would also like to thank the reviewers for their feedback. This research is supported by NSF Award \#2219864, the Brown Institute for Media Innovation and the Stanford Institute for Human-Centered AI (HAI).
\end{acks}


\bibliographystyle{ACM-Reference-Format}
\bibliography{main}

\newpage
\appendix
\section{Appendix: Motion Graphics Video Test Set}
\label{sec:appendix}

We created a test set of 38 motion graphics video to evaluate our motion vectorization pipeline (Table~\ref{tab:tracking_eval}).
Tracking foreground objects through occlusions (either between objects
or at the edge of the frame as on object entry/exit) and across fast
motions (which we define as moments when an object's bounding box in
frame $F_{t-1}$ does not overlap with its bounding box in $F_t$) is
especially challenging. Many of the test videos contain such challenging features.
%
A few of the more challenging videos also contain textures, photographic elements or color gradients in the
foreground or background (marked with $\ddagger$).
The two rightmost columns of Table~\ref{tab:tracking_eval} show the
 total number of tracking errors and a breakdown of these errors by
 mapping type; each element of the 8-tuple records the number of
 errors for corresponding mapping type as shown in
 Figure~\ref{fig:mappingTypes}. Thus, the video named {\em lucy}
 contains 4 one-to-one mapping errors (e.g., a region in $F_t$ is
 assigned an incorrect object ID) and 11 one-to-many (no split) errors
 (e.g., two or more regions that should be assigned the same object ID
 were incorrectly assigned different object IDs). See
 Figure~\ref{fig:eval} (bottom right) for an examples of this error for the {\em lucy} video.

 \begin{table}[h]
  \centering
      \caption{Our test set of 38 motion graphics videos. Six of the videos contain no occlusions and no fast motion. Twelve contain only occlusions and no fast motions. Seven contain only fast motion. Thirteen contain both. Some of the videos contain textures, photographic elements or color gradients in the foreground or background (marked with $\ddagger$). The reconstruction $L_2$ error shows the average RGB error for the SVG motion program produced by our vectorization pipeline. The rightmost columns show the total number tracking errors (all) and the errors by mapping type (Figure~\ref{fig:mappingTypes}).
    }
  \small{
    \begin{NiceTabular}{|l|r|r|r|r|r|}
    \hline
    \multirow{2}{*}{\textbf{Video}} & \multirow{1}{*}{\textbf{Num. }} & \multirow{1}{*}{\textbf{Num.}} & \multirow{1}{*}{\textbf{Recon.}} & \multicolumn{2}{c}{\textbf{Tracking errors}} \\
    \cline{5-6}
     & \multicolumn{1}{c}{\bf frames} & \multicolumn{1}{c}{\bf objs} & \multicolumn{1}{c}{\bf $L_2$ error} & \multicolumn{1}{c}{All} & \multicolumn{1}{c}{Mapping type} \\
    \hline
    \multicolumn{6}{|l|}{\bf No occlusions and no fast motion} \\
    \hline
    ball2  &           500 &              4 &   0.0034 &  0  &   0,0,0,0,0,0,0,0 \\
    \hline
    ball3  &           215 &              8 &    0.0024 &  0  &   0,0,0,0,0,0,0,0 \\
    \hline
    eyes  &           312 &             14 &   0.0050 &   0  &   0,0,0,0,0,0,0,0 \\
    \hline
    format  &           151 &              6 &    0.0036 &  0  &   0,0,0,0,0,0,0,0 \\
    \hline
    levers  &           144 &              6 &    0.0063 &  0  &   0,0,0,0,0,0,0,0 \\
    \hline
    support  &           299 &              9 &    0.0024 &  0  &   0,0,0,0,0,0,0,0 \\

    \hline
    \multicolumn{6}{|l|}{\bf Occlusions only} \\
    \hline
    dog  &           133 &             12 &    0.017  &  0  &   0,0,0,0,0,0,0,0 \\
    \hline
    five  &           144 &              5 &    0.0024 &   0  &   0,0,0,0,0,0,0,0 \\
    \hline
    giftbox1  &            80 &              8 &    0.0078 &  0  &   0,0,0,0,0,0,0,0 \\
    \hline
    giftbox2 &            80 &             10 &     0.012 &  0  &   0,0,0,0,0,0,0,0 \\
    \hline
    hype1  &           144 &              4 &     0.022 & 0  &   0,0,0,0,0,0,0,0 \\
    \hline
    hype2  &           144 &              4 &     0.024 & 0  &   0,0,0,0,0,0,0,0 \\
    \hline
    pingpong  &           144 &             21 &    0.0093 &  0  &   0,0,0,0,0,0,0,0 \\
    \hline
    playDesign  &           438 &             13 &    0.0068 &  0  &   0,0,0,0,0,0,0,0 \\
    \hline
    sundance &           336 &             70 &    0.0071 &  0  &   0,0,0,0,0,0,0,0 \\
    \hline
    ball5  &           289 &              4 &   0.0072 &  0  &   0,0,0,0,0,0,0,0 \\
    \hline
    sydney ($\ddagger$) &            98 &             44 &   0.0394 &   4  &   4,0,0,0,0,0,0,0 \\
    \hline
    morningShow   &           147 &            162 &     0.011 & 5  &   5,0,0,0,0,0,0,0 \\

    \hline
    \multicolumn{6}{|l|}{\bf Fast motion only} \\
    \hline
    ball4   &            79 &              2 &   0.0026 &   0  &   0,0,0,0,0,0,0,0 \\
    \hline
    book2 ($\ddagger$)  &            36 &             36 &   0.0095 &    0  &   0,0,0,0,0,0,0,0 \\
    \hline
    transforms   &           358 &             27 &    0.0034 &  0  &   0,0,0,0,0,0,0,0 \\
    \hline
    seesaw ($\ddagger$)   &           188 &              4 &    0.0017 &   0  &   0,0,0,0,0,0,0,0 \\
    \hline
    wordAWeek   &           151 &             12 &    0.0036 &   0  &  0,0,0,0,0,0,0,0 \\
    \hline
    deconstruct  &           156 &             11 &      0.0010 & 0  &   0,0,0,0,0,0,0,0 \\
    \hline
    beautiful   &           221 &             16 &    0.0037 &   5  &   4,1,0,0,0,0,0,0 \\

    \hline
    \multicolumn{6}{|l|}{\bf Both occlusions and fast motion} \\
    \hline
    ball1 ($\ddagger$)  &           394 &              2 &    0.0083 &  0  &   0,0,0,0,0,0,0,0 \\
    \hline
    face   &           156 &              5 &     0.0011 & 0  &   0,0,0,0,0,0,0,0 \\
    \hline
    filmRadio   &           177 &             60 &     0.0040 &  1  &   1,0,0,0,0,0,0,0 \\
    \hline
    183   &            96 &             32 &     0.010 &   2  &   2,0,0,0,0,0,0,0 \\
    \hline
    gsuite ($\ddagger$) &           481 &             24 &     0.017 &  3  &   3,0,0,0,0,0,0,0 \\
    \hline
    book1 ($\ddagger$) &           108 &              7 &    0.0036 &   4  &   4,0,0,0,0,0,0,0 \\
    \hline
    kapptivate   &            50 &             13 &   0.0063 &  4  &   3,0,0,0,0,0,0,1 \\
    \hline
    avokiddo   &           130 &             20 &     0.0033 & 6  &   6,0,0,0,0,0,0,0 \\
    \hline
    dates ($\ddagger$)  &            181 &             36 &   0.023 &   6  &   6,0,0,0,0,0,0,0 \\
    \hline    
    5k ($\ddagger$)  &           119 &             18 &   0.033 &   7  &   4,3,0,0,0,0,0,0 \\
    \hline
    shapeman   &            70 &             14 &    0.0048 &   10  &  6,3,1,0,0,0,0,0 \\
    \hline
    confetti   &            45 &             143 &  0.012 &   13  &   12,0,1,0,0,0,0,0 \\
    \hline
    lucy   &           353 &             33 &    0.013 &  15  &   4,11,0,0,0,0,0,0 \\
    \hline
    \end{NiceTabular}
    }
    \label{tab:tracking_eval}
\end{table}


\end{document}